# Models of topological barriers and molecular motors of bacterial DNA

Marc JOYEUX

*Laboratoire Interdisciplinaire de Physique, CNRS and Université Grenoble Alpes, Grenoble, France*

Contact : Marc JOYEUX. Email : marc.joyeux@univ-grenoble-alpes.fr

# Models of topological barriers and molecular motors of bacterial DNA

Bacterial genomes are partitioned into kilobases long domains that are topologically independent from each other, meaning that change of DNA superhelicity in one domain does not propagate to neighbours. This is made possible by proteins like the LacI repressor, which behave like topological barriers and block the diffusion of torsion along the DNA. Other proteins, like DNA gyrases and RNA polymerases, called molecular motors, use the energy released by the hydrolysis of ATP to apply forces and/or torques to the DNA and modify its superhelicity. Here, we report on simulation work aimed at enlightening the interplay between DNA supercoiling, topological barriers, and molecular motors. To this end, we developed a coarse-grained Hamiltonian model of topological barriers and a model of molecular motors and investigated their properties through Brownian dynamics simulations. We discuss their influence on the contact map of a model nucleoid and the steady state values of twist and writhe in the DNA. These coarse-grained models, which are able to predict the dynamics of plectonemes depending on the position of topological barriers and molecular motors, should prove helpful to back up experimental efforts, like the development of Chromosome Conformation Capture techniques, and decipher the organisational mechanisms of bacterial chromosomes.

**1. Introduction**

The chromosomal DNA of bacteria is folded into a compact body called the nucleoid [1,2], which is composed essentially of DNA ($\approx 80\%$), RNA ($\approx 10\%$), a dozen of major nucleoid proteins ($\approx 10\%$) [3], DNA polymerases, RNA polymerase, and about hundred species of transcription factors. In contrast with the nucleus of eukaryotic cells, the nucleoid is not enclosed in a membrane, but it still occupies only part of the cell volume [1,2]. The bacterial chromosome itself is organized into independent topological domains, whose torsional state is not affected by the torsional state of neighbouring domains [4,5]. For example, the genome of *E. coli* is composed of several hundred topological domains with average size $\approx 10000$ base pairs (bp) [6]. Organization of the DNA into independent domains is made possible by proteins, which behave like *topological barriers* and block the diffusion of torsional stress along the DNA. Proteins capable of acting as topological barriers include (i) actively transcribing RNA polymerases [7-9] and (ii) certain DNA-bridging nucleoid proteins, like pairs of LacI repressor [10-13], which form DNA loops that are topologically isolated from the rest of the chromosome [10,14-16]. In a recent work [17], we determined under which conditions DNA-bridging proteins may act as topological barriers. To this end, we developed a coarse-grained bead-and-spring model and investigated its properties through Brownian dynamics simulations. We showed that proteins must block the diffusion of the excess of twist through the two binding sites on the DNA molecule and, simultaneously, prevent the rotation of one DNA segment relative to the other one [17].

Other proteins called molecular motors introduce *activity* in the DNA coil, that is, they use the energy released by the hydrolysis of ATP to apply forces and/or torques to the DNA molecule. For example, the constitutive level of DNA tension is in the range from 0.1 to 1.0 pN [18], but bursts of significantly larger forces arise from the action of DNA polymerases and RNA polymerases [19-21]. Moreover, alteration of the torsional state of bacterial circular DNA is mediated by the recruitment of topoisomerases, which can either increase or decrease the winding of the double helix [22]. The maximal value of superhelical density generated by DNA gyrases and RNA polymerases is of the order of $\sigma = -0.12$ [23-25], that is, about 4 times the superhelical density of protein-bound DNA in living *E. coli* cells [26] and twice that of naked DNA *in vitro* [27].

In the present paper, we report on simulation work aimed at enlightening the interplay between DNA supercoiling, topological barriers, and active mechanisms. To this end, we revisited and extended the results of our recent work [17], where the role of activity was ignored, in spite of the fact that activity was (unwittingly) introduced at two different places. We first point out in Section 3.1 that a counter-intuitive result discussed in [17] was actually due to an erroneous symmetrisation in the expression of torsional forces, which introduced activity in the model and modified sensitively the final steady state reached by the system. We next propose in Section 3.2 a Hamiltonian model of topological barrier, which does not result in energy being injected into the DNA loop, in contrast with the model proposed in [17], and discuss the differences between the two models. Finally, we propose in Section 3.3 a model of molecular motor that exerts a constant torque on the DNA helix and study its effects on the contact map of the nucleoid. It is anticipated that simulations based on the models of topological barrier and molecular motor proposed in the present paper may help in understanding and deciphering experimental results, like those obtained with Chromosome Conformation Capture set-ups [28].

## 2. Simulation details and methods

### 2.1 DNA molecule

The model of DNA molecules used in this work was described in the Supporting Material of [17]. It is summarized here briefly for the sake of completeness and clarity.

Circular DNA molecules are modeled as circular chains of $n$ beads ($n = 600$ or $n = 2880$) with radius $a = 1.0$ nm separated at equilibrium by a distance $l_0 = 2.5$ nm. Each bead represents 7.5 base pairs (bp), so that chains of 600 and 2880 beads represent a plasmid of 4500 bp and a longer DNA molecule of 21600 bp, respectively. The potential energy of the DNA molecule, $E_{\text{DNA}}$, consists of four terms

$$E_{\text{DNA}} = V_s + V_b + V_t + V_e, \tag{1}$$

where

$$V_s = \frac{h}{2} \sum_{k=1}^{n} (l_k - l_0)^2, \tag{2}$$

$$V_b = \frac{g}{2} \sum_{k=1}^{n} \theta_k^2, \tag{3}$$

$$V_t = \frac{\tau}{2} \sum_{k=1}^{n} (\Phi_{k+1} - \Phi_k)^2, \tag{4}$$

$$V_e = q^2 \sum_{k=1}^{n-2} \sum_{K=k+2}^{n} H(\|\mathbf{r}_k - \mathbf{r}_K\| - 2a), \tag{5}$$

describe the stretching, bending, torsional, and electrostatic energy of the DNA chain, respectively. $V_s$, $V_b$ $V_e$ are expressed in terms of a set of vectors $\mathbf{r}_k$, which represent the position of bead $k$ in the space-fixed frame, with the convention that $\mathbf{r}_{k+n} \equiv \mathbf{r}_k$. In Equation (2), $l_k = \|\mathbf{r}_{k+1} - \mathbf{r}_k\|$ denotes the distance between successive beads $k$ and $k+1$.

In Equation (3), $\theta_k$ denotes the angle between vectors $\mathbf{r}_{k+1} - \mathbf{r}_k$ and $\mathbf{r}_{k+2} - \mathbf{r}_{k+1}$. Torsional energy $V_t$ is instead expressed in terms of the rotation of body-fixed frames $(\mathbf{u}_k, \mathbf{f}_k, \mathbf{v}_k)$, which are attached to each bead $k$. $\mathbf{u}_k = (\mathbf{r}_{k+1} - \mathbf{r}_k)/l_k$ denotes the unit vector pointing from bead $k$ to bead $k+1$, whereas $\mathbf{f}_k$ and $\mathbf{v}_k$ describe the rotation of the DNA helix perpendicular to $\mathbf{u}_k$. For example, it may be considered that $\mathbf{f}_k$ points towards the major groove. Angle $\Phi_k$ quantifies the rotation around $\mathbf{u}_k$ of $(\mathbf{f}_{k+1}, \mathbf{v}_{k+1})$ with respect to $(\mathbf{f}_k, \mathbf{v}_k)$, with the convention that $\Phi_{k+n} \equiv \Phi_k$.

The stretching rigidity in Equation (2) was set to $h = 100\, k_B T / l_0^2$, where $T = 295$ K, because this value ensures that the variations of the distance between successive beads remain small enough when reasonably large integration time steps are used [29]. The value of $h$ is about 4 orders of magnitude smaller than real stretching force constants of the DNA backbone. In contrast, the bending rigidity in Equation (3) was deduced from the known persistence length of DNA, $\xi = 50$ nm, according to $g = \xi k_B T / l_0 = 20\, k_B T$. The torsional contribution $V_t$ in Equation (4) was borrowed from Reference [30] and torsion forces and momenta were computed as described therein, except where otherwise stated. The value of the torsional rigidity, $\tau = 25\, k_B T$, ensures that the writhe contribution $Wr$ accounts for approximately 2/3 of the linking number difference $\Delta Lk$ at equilibrium [31], see Figure S1 of [17]. Finally, the electrostatic energy of the DNA chain is expressed as a sum of repulsive Debye-Hückel terms with hard core. Function $H(r)$ in Equation (5) is defined according to

$$H(r) = \frac{1}{4\pi\varepsilon r}\exp\left(-\frac{r}{r_D}\right), \tag{6}$$

where $\varepsilon = 80\, \varepsilon_0$ denotes the dielectric constant of the buffer and $r_D = 1.07$ nm the Debye length inside the buffer. This value of the Debye length corresponds to a concentration of monovalent salt of 100 mM, which is the value that is generally assumed for the cytoplasm of bacterial cells. $q = -3.52\, \bar{e}$, where $\bar{e}$ is the absolute charge of the electron, is the value of the electric charge, which is placed at the centre of each DNA bead. This value was deduced from Manning's counterion condensation theory [17,32,33].

*2.2 Cell membrane*

The longer DNA chain with $n = 2880$ beads is enclosed in a confinement sphere of radius $R_0 = 120$ nm. The contour length of the DNA chain and the volume of the confinement sphere correspond approximately to 1/200th of their values in *E. coli* cells, so that the DNA base pair concentration of the model is close to the physiological value (about 10 mM). The confinement energy $E_{\text{wall}}$ is expressed in the form

$$E_{\text{wall}} = \zeta \sum_{k=1}^{n} f(\|\mathbf{r}_k\|), \quad (7)$$

where the repulsive force constant $\zeta$ was set to $1000 k_B T$ and the function $f(r)$ is defined according to

$$\begin{aligned} r \leq R_0 &: f(r) = 0 \\ r > R_0 &: f(r) = \left(\frac{r}{R_0}\right)^6 - 1 \end{aligned}, \quad (8)$$

*2.3 Langevin equations*

The dynamics of the system was investigated by integrating numerically overdamped Langevin equations. Practically, the updated positions and torsion angles at time step *i*+1 were computed from the positions and torsion angles at time step *i* according to

$$\mathbf{r}_k^{(i+1)} = \mathbf{r}_k^{(i)} + \frac{\Delta t}{6\pi\eta a} \mathbf{f}_k^{(i)} + \sqrt{\frac{2 k_B T \, \Delta t}{6\pi\eta a}} \, \mathbf{x}_k^{(i)}, \quad (9)$$

$$\Phi_k^{(i+1)} = \Phi_k^{(i)} + \frac{\tau \Delta t}{4\pi\eta a^2 l_0} (\Phi_{k+1}^{(i)} - 2\Phi_k^{(i)} + \Phi_{k-1}^{(i)}) + \sqrt{\frac{2 k_B T \, \Delta t}{4\pi\eta a^2 l_0}} \, X_k^{(i)}, \quad (10)$$

where the $\mathbf{f}_k^{(i)}$ are vectors of inter-particle forces arising from the total potential energy of the system, $T = 298$ K is the temperature of the system, $\mathbf{x}_k^{(i)}$ and $X_k^{(i)}$ are vectors of random numbers extracted from a Gaussian distribution of mean 0 and variance 1, $\eta = 0.00089$ Pa s is the viscosity of the buffer at 298 K, and $\Delta t = 10$ ps is the integration time step.

*2.4 Topological barrier*

In [17], we determined under which conditions a protein that bridges beads $\alpha$ and $\beta$ of a circular DNA chain may work as a topological barrier [10-13]. More precisely, the goal was to determine which constraints the protein must exert on these beads in order to separate the DNA chain into two topologically independent loops. A diagram of the DNA chain in the neighbourhood of beads $\alpha$ and $\beta$ is shown in Figure 1. The DNA chain is formally divided into two segments, namely segment $S=1$, which extends from bead $\alpha$ to bead $\beta$ and is shown in green, and segment $S=2$, which extends from bead $\beta$ to bead $\alpha$ (through beads $n$ and 1) and is shown in red. It was found in [17] that the protein must block the diffusion of the excess of twist through beads $\alpha$ and $\beta$ and, simultaneously, prevent the rotation of segment $(\alpha-1, \alpha, \alpha+1)$ relative to segment $(\beta-1, \beta, \beta+1)$. This was achieved in [17] by introducing a potential energy term $E_{\text{BP}}$, which mimics the action of the protein on the two binding sites,

$$E_{\text{BP}} = \frac{h}{2}(d_{\alpha\beta} - d_{\alpha\beta}^0)^2 + \frac{5g}{2}(\xi_\alpha - \frac{\pi}{2})^2 + \frac{5g}{2}(\xi_\beta - \frac{\pi}{2})^2 + \frac{5g}{2}(\psi_{\alpha\beta} - \frac{2\pi}{3})^2, \quad (11)$$

and imposing that $\mathbf{f}_\alpha$ and $\mathbf{f}_\beta$ remain perpendicular to $\mathbf{r}_{\alpha\beta} = \mathbf{r}_\beta - \mathbf{r}_\alpha$ at all times, that is

$$\mathbf{f}_\alpha \cdot \mathbf{r}_{\alpha\beta} = \mathbf{f}_\beta \cdot \mathbf{r}_{\alpha\beta} = 0. \quad (12)$$

In Equation (11), $d_{\alpha\beta}$ denotes the distance between the centres of beads $\alpha$ and $\beta$. The value $d_{\alpha\beta}^0 = 4$ nm is small enough to ensure that no DNA segment can cross the line between $\alpha$ and $\beta$, which is formally occupied by the DNA bridging protein [17]. In the same equation, $\psi_{\alpha\beta}$ denotes the angle between vectors $\mathbf{u}_\alpha$ and $\mathbf{u}_\beta$ (see Figure 1), so that the last term in the right-hand side of Equation (11) restricts strongly the amplitude of the rotation of the DNA segment containing bead $\alpha$ with respect to the DNA segment containing bead $\beta$.

In [17], the requirement of Equation (12) was fulfilled by computing, after each integration step, the angles

$$\delta_k = -\tan^{-1}(\frac{\mathbf{f}_k \cdot \mathbf{r}_{\alpha\beta}}{\mathbf{v}_k \cdot \mathbf{r}_{\alpha\beta}}) \quad (13)$$

for $k = \alpha, \beta$, and rotating $\mathbf{f}_k$ and $\mathbf{v}_k$ around $\mathbf{u}_k$ by $\delta_k$

$$\begin{aligned}\mathbf{f}_k &\to \cos\delta_k\, \mathbf{f}_k + \sin\delta_k\, \mathbf{v}_k \\ \mathbf{v}_k &\to -\sin\delta_k\, \mathbf{f}_k + \cos\delta_k\, \mathbf{v}_k\end{aligned} \qquad (14)$$

for $k = \alpha, \beta$. The angles $\delta_k$ in Equation (13) and the corrections in Equation (14) remain small if $\mathbf{r}_{\alpha\beta}$ is not perpendicular simultaneously to $\mathbf{f}_k$ and $\mathbf{v}_k$, that is, if vectors $\mathbf{r}_{\alpha\beta}$ and $\mathbf{u}_k$ ($k = \alpha, \beta$) are not collinear. It is precisely the role of the second and third term in the right-hand side of Equation (11) to prevent such collinearity. $\xi_\alpha$ denotes the angle between vectors $-\mathbf{r}_{\alpha\beta}$ and $\mathbf{u}_\alpha$ and $\xi_\beta$ the angle between vectors $\mathbf{r}_{\alpha\beta}$ and $\mathbf{u}_\beta$ (see Figure 1). These two terms consequently ensure that $\xi_\alpha$ and $\xi_\beta$ deviate only moderately from $\pi/2$, so that $\mathbf{r}_{\alpha\beta}$ and the $\mathbf{u}_k$ remain far from collinearity.

## *2.5 Computation of the excess of twist and the writhe*

The excess of twist $\Delta T\mathrm{w}^{(S)}$, the writhe $W\mathrm{r}^{(S)}$ and the linking number difference $\Delta L\mathrm{k}^{(S)}$ are computed for each segment $S$ comprised between beads $k_{\min}$ and $k_{\max}$ according to

$$\Delta T\mathrm{w}^{(S)} = \frac{1}{2\pi} \sum_{k=k_{\min}}^{k_{\max}} (\Phi_{k+1} - \Phi_k), \qquad (15)$$

$$W\mathrm{r}^{(S)} = \frac{1}{2\pi} \sum_{k=k_{\min}}^{k_{\max}} \sum_{K=k+1}^{k_{\max}} \frac{[(\mathbf{r}_{k+1}-\mathbf{r}_k)\times(\mathbf{r}_{K+1}-\mathbf{r}_K)]\cdot(\mathbf{r}_k-\mathbf{r}_K)}{|\mathbf{r}_k - \mathbf{r}_K|^3}, \qquad (16)$$

$$\Delta L\mathrm{k}^{(S)} = \Delta T\mathrm{w}^{(S)} + W\mathrm{r}^{(S)}, \qquad (17)$$

whereas the inter-segment writhe $W\mathrm{r}^{(1,2)}$ resulting from the entanglement of the two segments is computed according to

$$W\mathrm{r}^{(1,2)} = \frac{1}{2\pi} \sum_{k=\alpha}^{\beta} \sum_{K=\beta}^{\alpha+n} \frac{[(\mathbf{r}_{k+1}-\mathbf{r}_k)\times(\mathbf{r}_{K+1}-\mathbf{r}_K)]\cdot(\mathbf{r}_k-\mathbf{r}_K)}{|\mathbf{r}_k - \mathbf{r}_K|^3} \qquad (18)$$

The total excess of twist $\Delta T\mathrm{w}$, the total writhe $W\mathrm{r}$, the total linking number difference $\Delta L\mathrm{k}$, and the superhelical density $\sigma$ of the DNA chain are subsequently obtained from

$$\Delta T\text{w}=\Delta T\text{w}^{(1)} +\Delta T\text{w}^{(2)} \tag{19}$$

$$Wr = Wr^{(1)} + Wr^{(2)} + Wr^{(1,2)} \tag{20}$$

$$\Delta L\text{k}=\Delta T\text{w}+Wr=\Delta L\text{k}^{(1)} +\Delta L\text{k}^{(2)} +Wr^{(1,2)} \tag{21}$$

$$\sigma = \frac{10.5\,\Delta L\text{k}}{7.5\,n} \tag{22}$$

## 3. Results and discussion

### *3.1 Effect of asymmetric torsional forces*

In order for the two loops extending on both sides of beads $\alpha$ and $\beta$ (segments 1 and 2) to be topologically independent, the protein that bridges beads $\alpha$ and $\beta$ must block the diffusion of $\Delta T\text{w}$ through the two beads and prevent the rotation of the segment centred on $\alpha$ relative to the segment centred on $\beta$ [17]. If the second condition is not fulfilled, that is, if the effect of the DNA-bridging protein is modelled by Equations (12)-(14) and

$$E_{\text{BP}} = \frac{h}{2}(d_{\alpha\beta} - d_{\alpha\beta}^0)^2 + \frac{5g}{2}(\xi_\alpha - \frac{\pi}{2})^2 + \frac{5g}{2}(\xi_\beta - \frac{\pi}{2})^2 \tag{23}$$

instead of Equation (11), then $\Delta L\text{k}^{(2)} - \Delta L\text{k}^{(1)}$ remains constant along the trajectory, but $\Delta L\text{k}^{(1)}$, $\Delta L\text{k}^{(2)}$ and $Wr^{(1,2)}$ do vary, which indicates that the two loops are not topologically independent from each other (Figure 4 of [17]).

More surprisingly, it was additionally found in [17] that, if the second condition is not fulfilled, then at long times $Wr^{(1)} + Wr^{(2)} \to 0$ (or equivalently $Wr \to Wr^{(1,2)}$), meaning that the entanglement of the two loops accounts for all the writhe of the DNA chain (Figure 4b of [17]). Very compact conformations with highly entangled DNA loops were accordingly observed (Figure 3b of [17]). We were unable to propose a rationale for this latter result [17]. Recently, we however noticed that a term was inappropriately symmetrised in the code used in [17] and that an error was consequently introduced in the computation of torsional forces. More precisely, according to Equations (A26) and (A27) of [30], the component of the torsional force acting on bead $i$ must be computed according to

$$\frac{\mathbf{F}_{ti}}{\tau} = (F_i + F_i^+)\mathbf{f}_i + (V_i + V_i^+)\mathbf{v}_i - (F_{i-1} + F_{i-1}^+)\mathbf{f}_{i-1} - (V_{i-1} + V_{i-1}^+)\mathbf{v}_{i-1}, \qquad (24)$$

whereas we mistakenly used

$$\frac{\mathbf{F}_{ti}}{\tau} = (F_i + V_{i+1}^+)\mathbf{f}_i + (V_i + V_{i+1}^+)\mathbf{v}_i - (F_{i-1} + F_i^+)\mathbf{f}_{i-1} - (V_{i-1} + V_i^+)\mathbf{v}_{i-1} \qquad (25)$$

(see [30] for the meaning of the various symbols). If Equation (25) is used instead of Equation (24), then the sum of all torsional forces exerted on the DNA chain is still zero, but the force exerted by bead $i$ on bead $i+1$ is no longer equal to minus the force exerted by bead $i+1$ on bead $i$. We checked that the net result of this asymmetry is that plectonemes travel at nearly constant speed along the DNA chain. This is clearly seen in Movie 1 of the Supplemental Material, which shows the time evolution of the contact map of the DNA chain with $n = 2880$ beads enclosed in the confinement sphere and submitted to a torsional constraint $\Delta Lk = -132$ ($\sigma = -0.064$), when no protein bridges beads $\alpha$ and $\beta$ and Equation (25) is used instead of Equation (24). It was considered that two DNA beads are in contact if their centres are separated by less than 10 nm. Each frame of Movie 1 shows the probability density for beads $i$ (abscissa) and $j$ (ordinate) to be in contact, and the movie shows the evolution of this map over a time window of 20 ms. Lines of increased contact probability perpendicular to the main diagonal of the map denote plectonemes. Movie 1 indicates that the plectonemes travel (slither) through the whole DNA chain in about 20 ms, that is, at a speed of roughly 1000 bp/ms. In contrast, if (correct) Equation (24) is used instead of (incorrect) Equation (25), then plectonemes fluctuate around their current position for long times, as can be checked in Movie 2 of the Supplemental Material, which shows the evolution of the contact map over a time window of 40 ms when Equation 24 is used.

  Breaking the symmetry of torsional forces by using Equation (25) instead of Equation (24) clearly introduces activity in the model, which manifests itself through the slithering of plectonemes along naked DNA chains. Moreover, asymmetry of torsional forces is also responsible for the puzzling limit $Wr \to Wr^{(1,2)}$ reported in [17]. This point was ascertained by running simulations using the correct expression for torsional forces in Equation (24) and involving DNA-bridging proteins described by Equations (12)-(14) and (23). These simulations confirm the result of [17] that such proteins fail to divide the DNA chain into two topologically independent loops, because $\Delta Lk^{(1)}$ and $\Delta Lk^{(2)}$ do not remain constant along the trajectories, in spite of the fact that $\Delta Lk^{(2)} - \Delta Lk^{(1)}$ does. However, the puzzling limit $Wr \to Wr^{(1,2)}$ of [17] no longer holds true. This is illustrated in Figure 2, which shows the result of a simulation performed with the DNA chain with $n=600$ beads submitted to a torsional constraint $\Delta Lk \approx -7$ ($\sigma \approx -0.016$). At time $t=0$, when the bond between beads $\alpha$ and $\beta$ is formed and the diffusion of twist through the two beads is blocked, the torsional constraint is essentially localized in segment 2 ($\Delta Lk^{(2)} \approx -7$ against $\Delta Lk^{(1)} \approx 1$). Figure 2 indicates that $\Delta Lk^{(1)}$, $\Delta Lk^{(2)}$, $Wr^{(1)}$, $Wr^{(2)}$ and $Wr^{(1,2)}$ fluctuate only moderately around their initial values. In particular, the inter-segment writhe $Wr^{(1,2)}$ remains small compared to the intra-segment write $Wr^{(1)} + Wr^{(2)}$, so that $Wr \approx Wr^{(1)} + Wr^{(2)}$, in strong contrast with the results in [17]. Moreover, the amplitude and the frequency of occurrence of the fluctuations of $\Delta Lk^{(1)}$, $\Delta Lk^{(2)}$ and $Wr^{(1,2)}$ away from their initial values are larger for moderate values of $|\Delta Lk|$ than for larger values of $|\Delta Lk|$, the reason being that tight plectonemes are quite efficient in blocking the rotation of the DNA segment containing bead $\alpha$ with respect to the segment containing bead $\beta$ and in preventing the entanglement of the two loops.

*3.2 Hamiltonian model of topological barrier*

Simulations involving DNA-bridging proteins acting as topological barriers (Equations (11)-(14)) and using the correct expression for torsional forces (Equation (24)) still display an unexpected feature that deserves closer scrutiny. This feature is illustrated in Figure 3a, which shows a snapshot extracted from a simulation performed with the DNA chain with $n=600$ beads submitted to a torsional constraint $\Delta Lk \approx -23$ ($\sigma \approx -0.054$). At time $t=0$, the torsional constraint is localized in segment 2 ($\Delta Lk^{(2)} \approx -23$ against $\Delta Lk^{(1)} \approx 0$). Because of the DNA-bridging protein acting as a topological barrier, $\Delta Lk^{(1)}$ and $\Delta Lk^{(2)}$ do not vary along the trajectory, in spite of the strong unbalance in the linking number differences. Segment 1 (in green in Figure 3a) nevertheless displays a rather complex geometry, which is quite unexpected owing to the fact that $\Delta Lk^{(1)} \approx 0$. Closer examination of the data indicates that the reason for this complex geometry is that $\Delta Tw^{(1)}$ evolves in a few tens of microseconds from $\approx 0$ to $\approx 2$ and $Wr^{(1)}$ from $\approx 0$ to $\approx -2$, while preserving a vanishing sum ($\Delta Tw^{(1)} + Wr^{(1)} = \Delta Lk^{(1)} \approx 0$). After the rapid initial evolution, $\Delta Tw^{(1)}$ and $Wr^{(1)}$ oscillate around nearly opposite values. In contrast, $\Delta T\text{w}^{(2)}$ remains close to $-8$ and $Wr^{(2)}$ close to $-15$ throughout the trajectory. At this point, it is worth remembering that, at thermodynamic equilibrium, the excess of twist and the writhe account for about one third and two thirds of the linking number difference, respectively [17,31]. In Figure 3a, segment 2 (red) is therefore at (or close to) equilibrium, whereas segment 1 (green) is instead far from equilibrium. All the simulations we performed display this non-equilibrium feature for the segment with $\Delta L\text{k}^{(S)} \approx 0$, the effect being all the more pronounced for larger values of $\left| \Delta L\text{k}^{(2)} - \Delta L\text{k}^{(1)} \right|$.

Non-equilibrium in the segment with $\Delta L\text{k}^{(S)} \approx 0$ is due to the fact that the condition in Equation (12) and its fulfilment through Equations (13) and (14) are non-Hamiltonian and therefore do not warrant energy conservation. The second and third terms in Equation (11) ensure that the correction introduced in the system by the use of Equations (13) and (14) is small at each time step, but do not provide any warranty concerning the evolution over long time intervals and many time steps. We checked that the two segments remain at thermodynamic equilibrium if $\Delta L\text{k}^{(1)} = \Delta L\text{k}^{(2)}$, which means that the effects of successive corrections cancel each other. However, for non-vanishing values of $\left| \Delta L\text{k}^{(2)} - \Delta L\text{k}^{(1)} \right|$, the effects of successive corrections clearly add up and maintain the segment with $\Delta L\text{k}^{(S)} \approx 0$ out of thermodynamic equilibrium. Stated in other words, Equations (12)-(14) describe an active process with injection of energy in the system.

Some proteins that are candidates for stabilizing supercoiled domains in bacterial chromosomes, like the *E. coli* and *Salmonella* MukBEF complex [34], are molecular motors which hydrolyse ATP [35]. It has been suggested that the energy generated by the MukBEF ATPase unit is used, like in ABC transporters [36] and the SMC-like protein Rad50 [37], to cause a massive transversal motion within the protein complex, for example by tilting the coiled arms of the complex [34]. It is probable that this motion breaks the thermodynamic equilibrium, as does the injection of energy in the model described by Equations (11)-(14). At present, there is of course no indication that Equations (11)-(14) may describe the MukBEF complex adequately and further work would be needed to model such energy-consuming topological barriers. Still, the model of Equations (11)-(14) highlights the interesting possibility that active topological barriers, in addition to isolating domains with different superhelicity, may also modify their contact maps by breaking the thermodynamic equilibrium, thereby producing for example more entangled structures.

Other proteins, like pairs of LacI repressors [10-13], are however passive systems, which do not consume energy and cannot be described by Equations (11)-(14). A Hamiltonian model of topological barriers is therefore required to describe these proteins. In order to find one, we scrutinized several models that hopefully block the diffusion of twist. It turned out, that one of these models works perfectly as a topological barrier and is nonetheless quite simple. According to this model, the effect of the protein bridging beads $\alpha$ and $\beta$ is described by

$$E_{\mathrm{BP}} = K(\mathbf{f}_\alpha . \mathbf{f}_\beta + \mathbf{v}_\alpha . \mathbf{v}_\beta), \tag{26}$$

meaning that the protein strives to maintain vectors $\mathbf{f}_\alpha$ and $\mathbf{f}_\beta$ (as well as vectors $\mathbf{v}_\alpha$ and $\mathbf{v}_\beta$) anti-parallel. This mechanism is sufficient to block the diffusion of twist through beads $\alpha$ and $\beta$ and replaces the non-Hamiltonian constraint in Equations (12)-(14). What was not clear *a priori* is whether the potential function in Equation (26) is also able to block the rotation of the two DNA segments relative to each other. We checked that this is indeed the case (see below). Setting $K$ to $125\, k_\mathrm{B}T$, that is 5 times the torsional rigidity $\tau$, worked fine for all the simulations we ran.

Using Equation (A21) of [30], according to which

$$\delta\mathbf{f}_k = -(\mathbf{f}_k . \delta\mathbf{u}_k)\mathbf{u}_k + (\delta\Phi_k)\mathbf{v}_k, \tag{27}$$

$$\delta\mathbf{v}_k = -(\mathbf{v}_k . \delta\mathbf{u}_k)\mathbf{u}_k - (\delta\Phi_k)\mathbf{f}_k, \tag{28}$$

the contributions of $E_\mathrm{BP}$ to the torque $T_k$ and force $\mathbf{F}_k$ felt by bead $k$ write

$$T_\alpha :+ K(\mathbf{f}_\alpha . \mathbf{v}_\beta - \mathbf{v}_\alpha . \mathbf{f}_\beta), \tag{29}$$

$$T_\beta :- K(\mathbf{f}_\alpha . \mathbf{v}_\beta - \mathbf{v}_\alpha . \mathbf{f}_\beta), \tag{30}$$

$$\mathbf{F}_\alpha :- \frac{K}{l_\alpha}(\mathbf{u}_\alpha . \mathbf{f}_\beta)\mathbf{f}_\alpha - \frac{K}{l_\alpha}(\mathbf{u}_\alpha . \mathbf{v}_\beta)\mathbf{v}_\alpha, \tag{31}$$

$$\mathbf{F}_{\alpha+1} := +\frac{K}{l_\alpha}(\mathbf{u}_\alpha.\mathbf{f}_\beta)\mathbf{f}_\alpha + \frac{K}{l_\alpha}(\mathbf{u}_\alpha.\mathbf{v}_\beta)\mathbf{v}_\alpha, \tag{32}$$

$$\mathbf{F}_\beta := -\frac{K}{l_\beta}(\mathbf{f}_\alpha.\mathbf{u}_\beta)\mathbf{f}_\beta - \frac{K}{l_\beta}(\mathbf{v}_\alpha.\mathbf{u}_\beta)\mathbf{v}_\beta, \tag{33}$$

$$\mathbf{F}_{\beta+1} := +\frac{K}{l_\beta}(\mathbf{f}_\alpha.\mathbf{u}_\beta)\mathbf{f}_\beta + \frac{K}{l_\beta}(\mathbf{v}_\alpha.\mathbf{u}_\beta)\mathbf{v}_\beta, \tag{34}$$

Figure 4 shows the result of a simulation performed with the same initial conditions as Figure 2, but with the DNA-bridging protein being described by Equation (26) instead of Equations (12)-(14) and (23). It is seen in this figure that $\Delta L\mathrm{k}^{(1)}$, $\Delta L\mathrm{k}^{(2)}$ and $Wr^{(1,2)}$ remain constant along the trajectory, meaning that Equation (26) describes an effective topological barrier, like Equations (11)-(14). However, unlike the model in Equations (11)-(14), both segments remain at thermodynamic equilibrium throughout the trajectory. This is illustrated in Figure 3b, which shows a snapshot extracted from a simulation with the same initial conditions as in Figure 3a, but with the DNA-bridging protein being modelled by Equation (26) instead of Equations (11)-(14). From the torsional point of view, segment 1 (green) looks much more relaxed in Figure 3b than in Figure 3a, and the simulation confirms that $Wr^{(1)} \approx 0.62 \Delta L\mathrm{k}^{(1)}$, as expected for a loop of this length at thermodynamic equilibrium. For comparison, we instead obtained $Wr^{(1)} \approx -6.70 \Delta L\mathrm{k}^{(1)}$ for the model in Equations (11)-(14).

Equation (26), together with the expression of torques and forces in Equations (27)-(34), consequently provides a simple but efficient model of DNA-bridging proteins acting as passive (Hamiltonian) topological barriers.

*3.3 Model of molecular motor*

It was shown in Section 3.1 that the use of Equation (25) for computing torsional forces introduces activity in the system, which results in plectonemes travelling monotonously along the DNA chain. Equation (25) can however not be considered as a model of molecular motor, because Equation (25) was applied to all the beads of the DNA chain, whereas a model of molecular motor should involve only the few beads to which the motor binds. A local model of molecular motor is described below and the resulting dynamics of the DNA chain is discussed.

Equation (25) of Reference [30], wherefrom the description of DNA torsion was borrowed, indicates that the vector $\mathbf{f}_k^{(i+1)}$ of the local basis at bead $k$ and time step $i+1$ must be computed from the local basis at bead $k$ and time step $i$ according to

$$\mathbf{f}_k^{(i+1)} = \mathbf{f}_k^{(i)} + (\Phi_k^{(i+1)} - \Phi_k^{(i)}) \mathbf{v}_k^{(i)} - (\mathbf{f}_k^{(i)} . \mathbf{u}_k^{(i+1)}) \mathbf{u}_k^{(i)}. \quad (35)$$

The component of $\mathbf{f}_k^{(i+1)}$ parallel to $\mathbf{u}_k^{(i+1)}$ is then deleted and $\mathbf{f}_k^{(i+1)}$ is renormalized. Vector $\mathbf{v}_k^{(i+1)}$ is finally obtained from the direct product

$$\mathbf{v}_k^{(i+1)} = \mathbf{u}_k^{(i+1)} \times \mathbf{f}_k^{(i+1)}. \quad (36)$$

Equations (35) and (36) describe the time evolution of the angular state of the slice of the DNA helix perpendicular to $\mathbf{u}_k$ at bead $k$. Modifying Equation (35) provides a simple way to model the action of molecular motors which modify the torsional state of the DNA helix. For example, the action of transcribing RNA-polymerases, which track the major groove of the DNA [38], thereby over-winding it downstream and under-winding it upstream [39], can be modelled by computing the new vector $\mathbf{f}_\gamma^{(i+1)}$ at bead $\gamma$ (where the RNA polymerase binds) according to

$$\mathbf{f}_\gamma^{(i+1)} = \mathbf{f}_\gamma^{(i)} + (\Phi_\gamma^{(i+1)} - \Phi_\gamma^{(i)} + \Delta\Phi) \mathbf{v}_\gamma^{(i)} - (\mathbf{f}_\gamma^{(i)} . \mathbf{u}_\gamma^{(i+1)}) \mathbf{u}_\gamma^{(i)}, \quad (37)$$

instead of Equation (35). Since the motion of most molecular motors along the DNA (around 40 bp/s for RNA polymerase [40]) is slow compared to the relaxation of twist (characteristic relaxation time of $\approx 1$ µs for this model [17]) and writhe (characteristic relaxation time of $\approx 200$ µs for this model [17])), $\gamma$ can be taken as a constant for most purposes, although it may be assumed to be a function of time whenever needed. The molecular motor described by Equation (37) exerts a torque of vector $\mathbf{u}_\gamma^{(i)}$ and magnitude $\tau \Delta\Phi$ at bead $\gamma$, which tends to increase continuously the twist on one side and decrease it on the other side, while leaving the linking number difference unaffected (see the bottom plot of Figure 5). In order to speed up calculations, simulations were run with $\Delta\Phi = 0.001$ rad, that is, a torque of about 100 pN.nm, which is roughly 5 to 10 times larger than the torque exerted by RNA polymerase on its substrate [41,42].

Movie 3 of the Supplemental Material shows the time evolution of the contact map of the DNA chain with $n=2880$ beads enclosed in the confinement sphere and submitted to a torsional constraint $\Delta Lk=-132$ ($\sigma=-0.064$) in the absence of any topological barrier, but with a molecular motor described by Equation (37) binding to bead $\gamma=720$. Movie 3 encompasses a total time window of 110 ms and the position of the molecular motor is indicated by green dashed lines. This movie indicates that the activity of the molecular motor triggers the reorganisation of the DNA from branched plectonemes to a single long plectoneme with the motor located close to one apex. Closer examination of the results reveals that the total number of supercoils remains nearly constant during the reorganisation of the plectonemes. Indeed, $Wr$ varies only slightly from $\approx$-92 to $\approx$-96 (see the middle plot of Figure 5), which is exactly compensated by a modest increase of the excess of twist from $\approx$-40 up to $\approx$-36 (see the top plot of Figure 5). Stated in other words, the continuous diffusion of twist generated by the molecular motor and the subsequent local interconversion from twist to writhe provoke a global reorganisation of the plectonemes but do not affect sensitively the ratio of writhe and twist.

Finally, Movie 4 of the Supplemental Material shows the time evolution over a total time window of 105 ms of the contact map of the DNA chain obtained from a simulation performed with both a topological barrier described by Equation (26) and a molecular motor described by Equation (37). As in Movie 3, the molecular motor binds to bead $\gamma=720$ of the DNA chain and its position is indicated by green dashed lines. The protein acting as a topological barrier bridges beads $\alpha=1$ and $\beta=1441$, the position of bead $\beta$ being indicated by red dashed lines. The red lines divide each frame of Movie 4 into four quadrants, the bottom left quadrant showing the contacts inside segment 1 (which contains the molecular motor), the top right quadrant the contacts inside segment 2, and the two other quadrants the contacts between segments 1 and 2. Clearly, the dynamics of the plectonemes in segment 2 resembles the dynamics of the plectonemes in Movie 2, meaning that the molecular motor in segment 1 does not affect the fluctuations of the plectonemes around their current position in segment 2. Conversely, the fact that beads $\alpha$ and $\beta$ block the diffusion of twist does not hinder the formation in segment 1 of a long plectoneme with the motor close to one apex, like in Movie 3. The passive (Hamiltonian) topological barrier described by Equation (26) is consequently able to separate a DNA chain into two loops with very different properties, namely a loop containing a molecular motor which promotes the formation of a single long plectoneme and a second loop containing shorter plectonemes which essentially fluctuate around their current position.

**4. Conclusion**

The introduction of Chromosome Conformation Capture (3C) techniques two decades ago [28,43] prompted an unprecedented effort to investigate DNA folding and understand its connections with chromosome functions. In particular, these techniques have shown that the genome of many species is organized into domains of preferential internal contacts, which are called "topologically associating domains" (TADs) [44]. TADs appear as a feature common to most organisms, although they display a great diversity in size, structure and mechanism of formation [44]. 3C techniques have evolved to the point where (a) cell-to-cell variability in the structure of the genome can be investigated [45,46], (b) contacts can be imaged at a resolution better than 1000 bp [47], and (c) measuring the time evolution of contact maps might be feasible shortly [48]. Coarse-grained models proved helpful to back up these revolutionary experimental efforts [49]. For example, the relative accuracy of three different experimental methods, Hi-C, GAM and SPRITE, has recently been questioned using the coarse-grained SBS model [50].

Studies performed in *C. crescentus* [51] and *B. subtilis* [52,53] have shown that the genome of prokaryotes is spatially organized into Chromosomal Interaction Domains (CIDs) ranging from 30 to 400 kbp, which resemble the TADs of eukaryotic cells. Moreover, regions enriched in plectonemes probably form CIDs, whereas boundaries between CIDs are essentially free of plectonemes [51]. Further studies performed in *M. pneumoniae* confirmed that the sharpness of CIDs depends on supercoiling [54]. It may therefore be safely anticipated, that coarse-grained models like those proposed in the present work, which are able to predict the dynamics of plectonemes depending on the position of topological barriers and molecular motors, will also prove helpful to decipher the organisational mechanisms of bacterial chromosomes.

**Declaration of interest**

The author reports there are no competing interests to declare.

**Figure captions**

Figure 1. Diagram showing the DNA chain in the neighbourhood of beads $\alpha$ and $\beta$ and illustrating the definition of the various symbols.

Figure 2. Time evolution of the writhe (top plot) and linking number difference (bottom plot) for a simulation performed with the DNA chain with 600 beads at $\Delta Lk = -7$ ( $\sigma = -0.016$ ) and a protein modelled by Equations (12)-(14) and (23).

Figure 3. Representative snapshots extracted from simulations performed with the DNA chain with 600 beads at $\Delta Lk \approx -23$ ( $\sigma \approx -0.054$ ) and a topological barrier modelled by (a) Equations (11)-(14), or (b) Equation (26).

Figure 4. Time evolution of the writhe (top plot) and linking number difference (bottom plot) for a simulation performed with the DNA chain with 600 beads at $\Delta Lk = -7$ ( $\sigma = -0.016$ ) and a topological barrier modelled by Equation (26).

Figure 5. Time evolution of the excess of twist (top plot), writhe (middle plot) and linking number difference (bottom plot) for a simulation performed with the DNA chain with 2880 beads at $\Delta Lk = -132$ ( $\sigma = -0.064$ ) when a molecular motor described by Equation (37) comes into operation at time $t=0$.

**FIGURE 1**

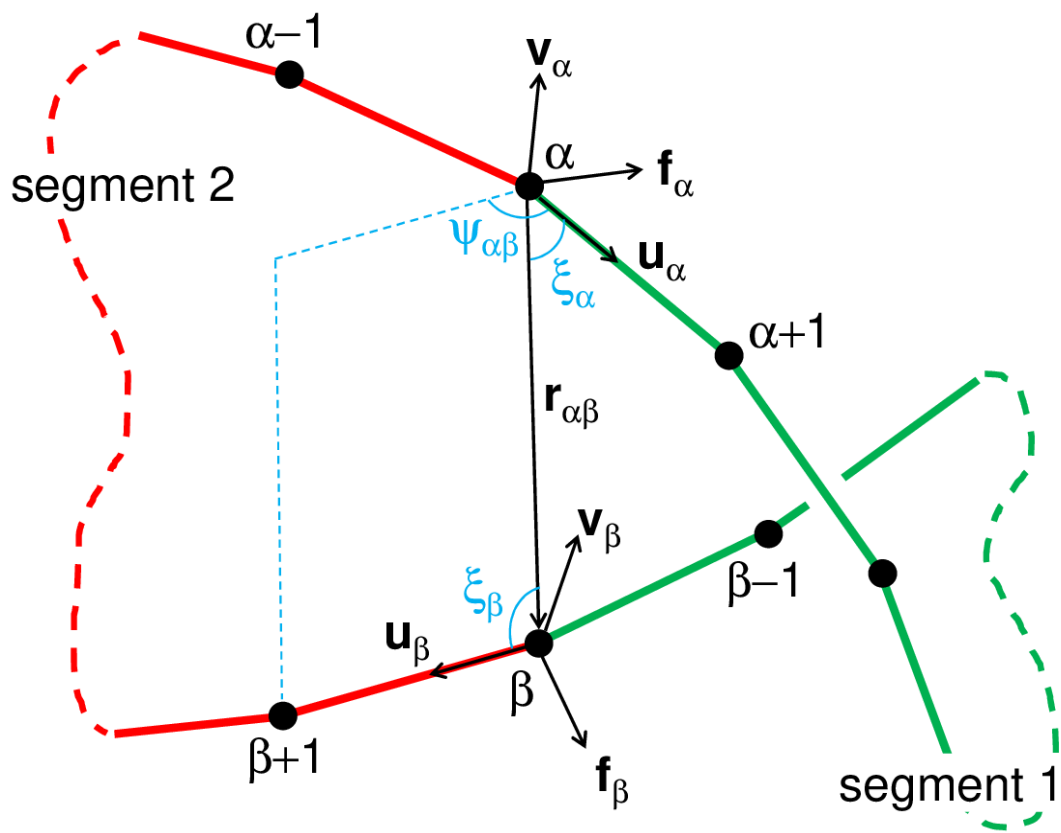

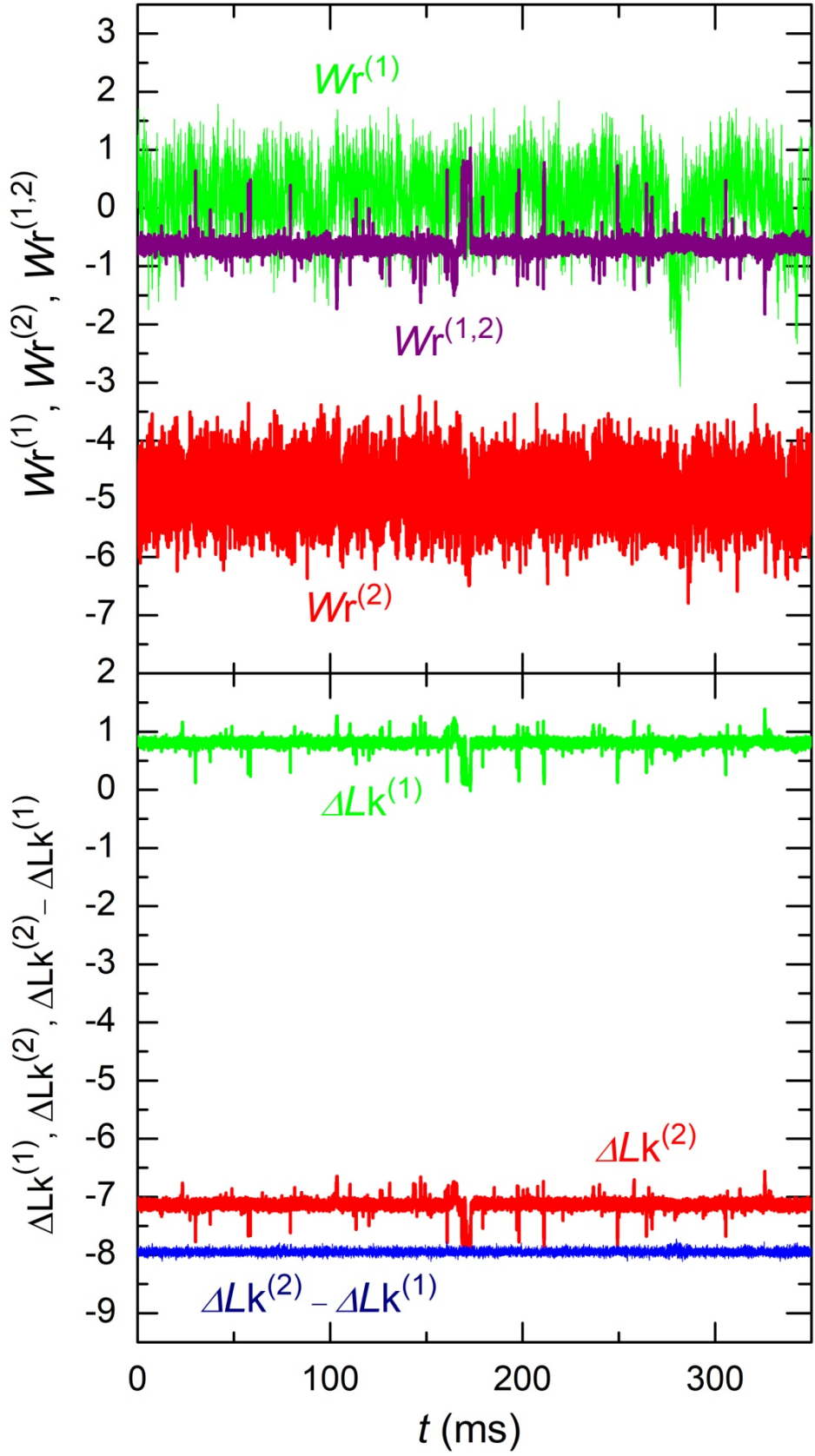

**FIGURE 3**

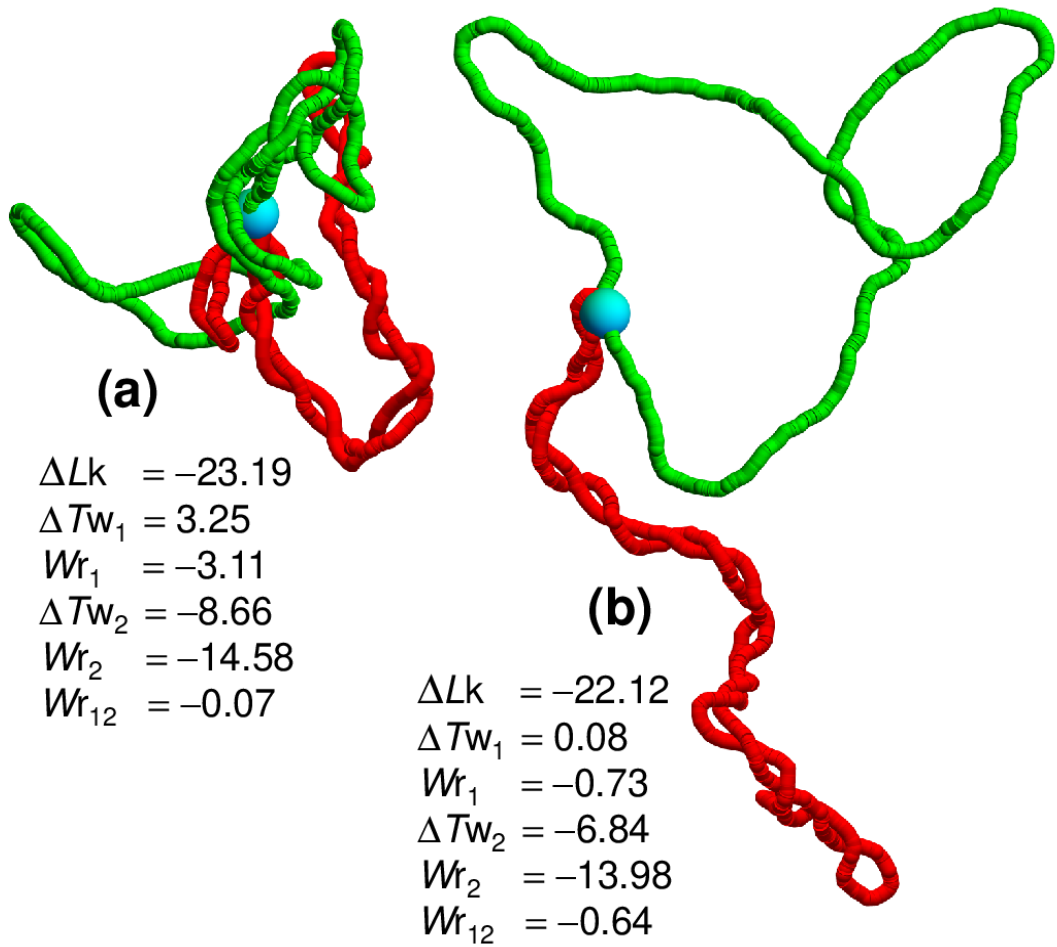

(a)
$\Delta L k = -23.19$
$\Delta T w_1 = 3.25$
$W r_1 = -3.11$
$\Delta T w_2 = -8.66$
$W r_2 = -14.58$
$W r_{12} = -0.07$

(b)
$\Delta L k = -22.12$
$\Delta T w_1 = 0.08$
$W r_1 = -0.73$
$\Delta T w_2 = -6.84$
$W r_2 = -13.98$
$W r_{12} = -0.64$

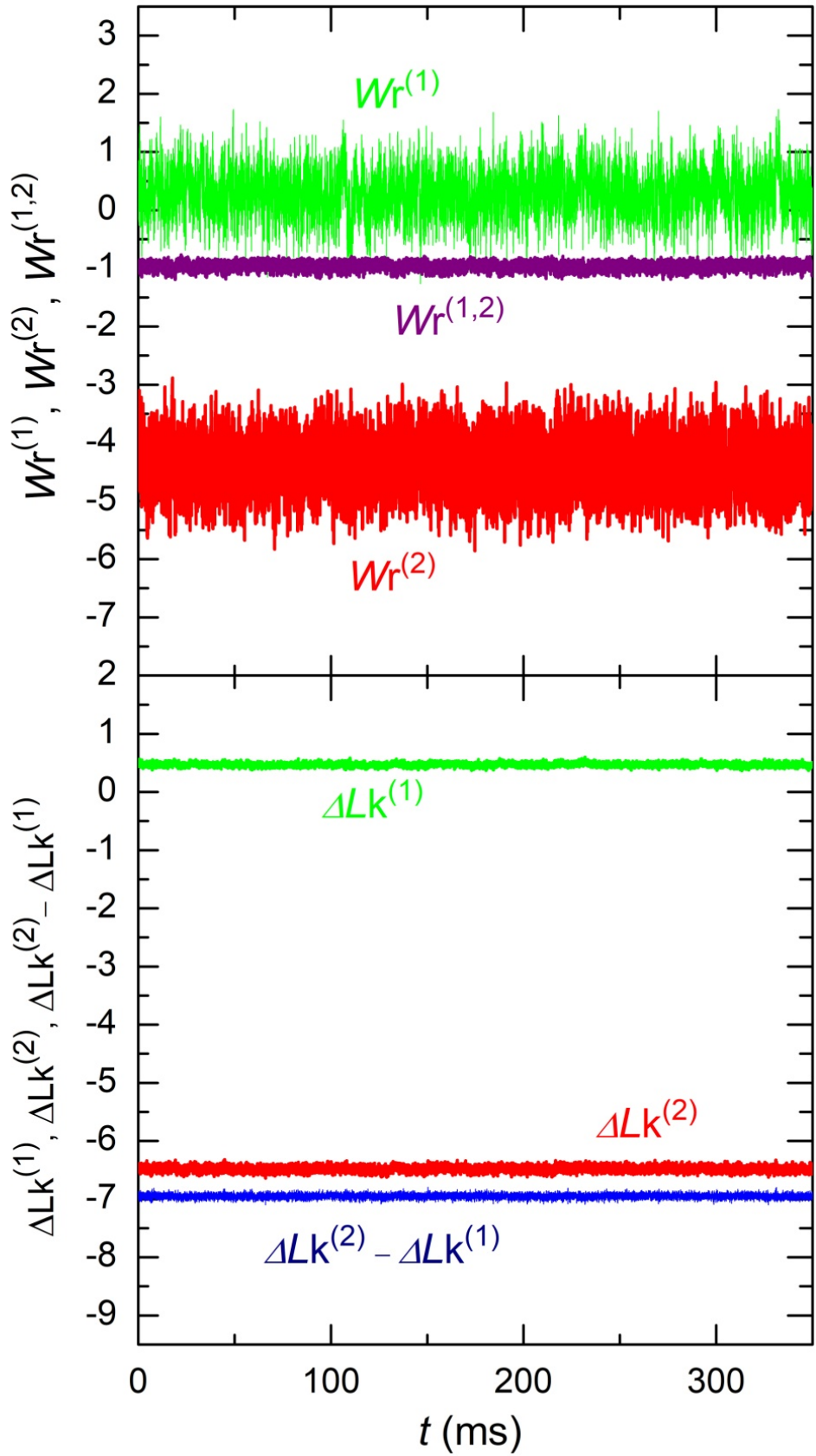

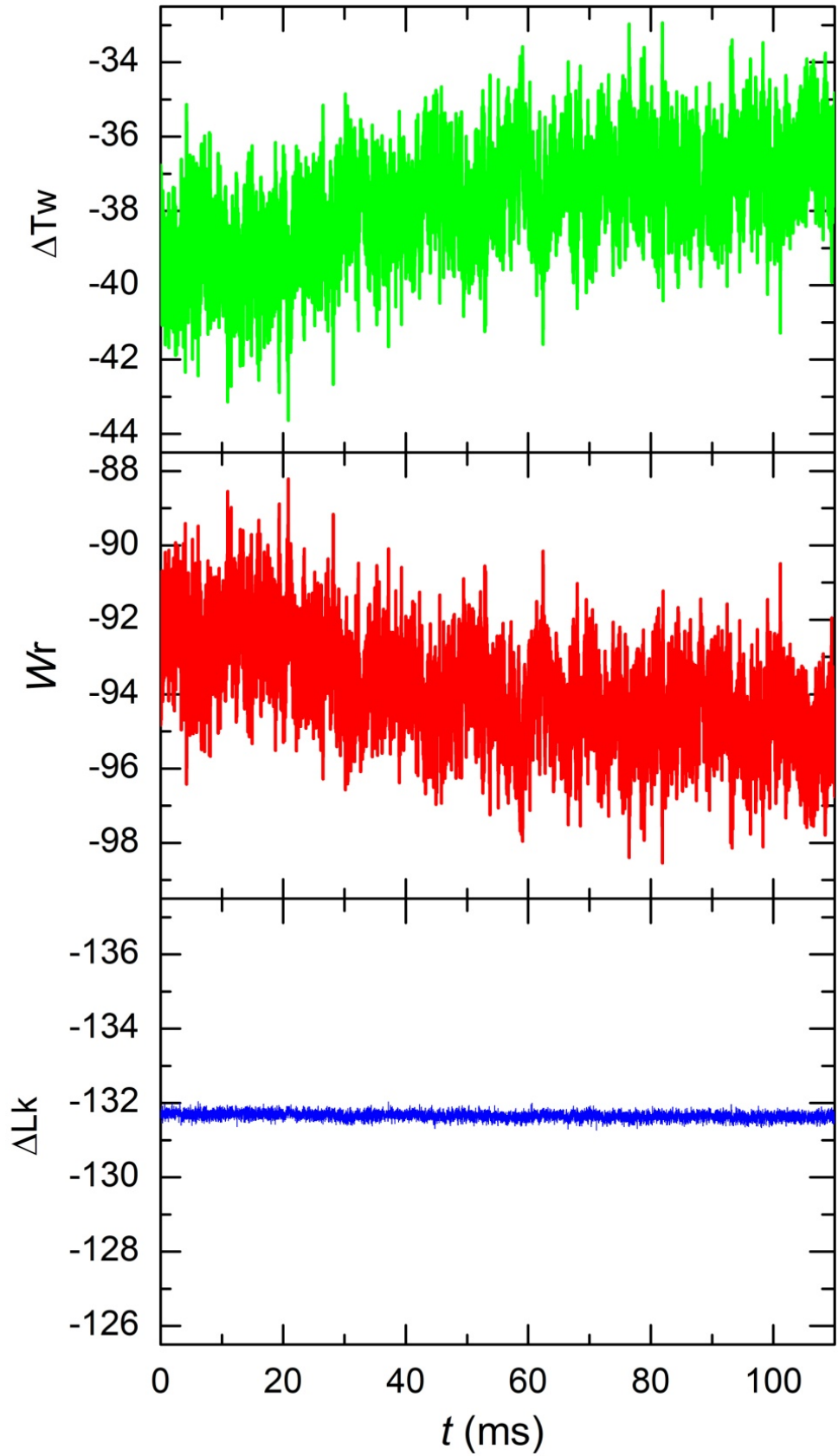